\documentclass[graybox]{svmult}

\usepackage{type1cm}        
\usepackage{makeidx}         
\usepackage{graphicx}        
\usepackage{multicol}        
\usepackage[bottom]{footmisc}

\usepackage{newtxtext}       %
\usepackage{newtxmath}       

\usepackage{todonotes}

\makeindex             

\begin{document}
	
	\title*{On Testing Data-Intensive Software Systems}
	\author{Michael Felderer, Barbara Russo and Florian Auer}
	\institute{Michael Felderer \at University of Innsbruck, Technikerstrasse 21a, 6020 Innsbruck, Austria
		\email{michael.felderer@uibk.ac.at}
		\and Barbara Russo \at Free University of Bozen-Bolzano, Piazza Universit\`a, 1, 39100 Bolzano, Italy
		\email{barbara.russo@unibz.it}
		\and Florian Auer \at University of Innsbruck, Technikerstrasse 21a, 6020 Innsbruck, Austria
		\email{florian.auer@uibk.ac.at}}
	%
	%
	\maketitle
	
	\abstract{Today's software systems like cyber-physical production systems or big data systems have to process large volumes and diverse types of data which heavily influences the quality of these so-called data-intensive systems. However, traditional software testing approaches rather focus on functional behavior than on data aspects. Therefore, the role of data in testing has to be rethought and specific testing approaches for data-intensive software systems are required. Thus, the aim of this chapter is to contribute to this area by (1) providing basic terminology and background on data-intensive software systems and their testing, and (2) presenting the state of the research and the hot topics in the area. Finally, the directions of research and the new frontiers on testing data-intensive software systems are discussed.}
	
	
	\section{Introduction}
	\label{sec:introduction}
	
	Information technology is evolving towards a kind of software-intensive systems that, more and more extensively, collect, integrate, process and analyze large data volumes that have to be processed with high velocity coming from a number of diverse data sources, so called "big data". Examples for such data-intensive systems can be found in all domains (e.g., finance, automotive, production) and all types of systems (e.g., information systems and cyber-physical systems). Especially also large and long-running cyber-physical systems that can be found in manufacturing plants are today data- and software-intensive as they are controlled by software and have to process large volumes and different types of data in real-time (see Section~\ref{sec:lcps}). The rapid growth of such systems is generating a paradigm shift in the field of software engineering. We are moving from traditional software engineering, which is functionality-centric, towards modern software and data engineering, which is data-centric and where the functionality is driven by the availability of data. Modern systems collect raw data from the users and their personal devices (like smart watches), from the environment and its smart objects (like sensors in industrial plants), as well as higher-level data coming from information providers like social platforms, open-data sites, and data silos. Also different types of data like image, sound, video or textual data have to be taken into account. Beyond functionality, the success (and added value) of such systems is tied to the availability of the data that is processed as well as its quality.
	
	Data-intensive (software) systems are therefore becoming increasingly prominent in today's world, where the collection, processing, and dissemination of ever-larger volumes of data has become a driving force behind innovation. However, data-intensive systems also pose new challenges to quality assurance~\cite{hummel2018collection} and especially testing. Testing software-intensive systems has traditionally focused on verifying and validating compliance and conformance to specifications, as well as some general nonfunctional requirements. Data-intensive systems require a different approach for testing and analysis, moving more towards exploring the system, its elements, behavior and properties from a big data and analytics perspective to help decision makers respond in real-time (e.g., for cyber-physical systems). The behavior and, therefore, the expected test result can often not be specified precisely, but only with a statistical uncertainty. For instance, the movement of a robot to assemble a part of a car can often not sufficiently be specified in an explicit way, but based on machine learning algorithms with uncertainty.
	
	As the area of data-intensive (software) systems and their testing is not well investigated so far, this chapter contributes by defining the main terminology and exploring existing approaches and promising future directions.
	
	The chapter is structured as follows. Section~\ref{sec:background} presents background on software testing and data quality. Section~\ref{sec:diss} defines what a data-intensive (software) system is and what main challenges to its quality assurance exist. Section~\ref{sec:lcps} discusses cyber-physical production systems as an example for data-intensive systems. Section~\ref{sec:testingdiss} presents an exploratory literature study on testing data-intensive software systems. Section~\ref{sec:discussion} discusses existing approaches to testing data-intensive systems and future directions. Finally, Section~\ref{sec:conclusion} concludes the chapter.

	\section{Background}
	\label{sec:background}
	
	In this section, we cover background on software testing and data quality.
	
	\subsection{Software Testing}
	\label{subsec:testing}
	
	Software testing (or software-intensive system testing) is a part of the overall engineering process of cyber-physical systems and can be defined as the process consisting of all lifecycle activities, both static and dynamic, concerned with planning, preparation and evaluation of software-intensive products, systems and services and related artifacts to determine that they satisfy specified requirements, to demonstrate that they are fit for purpose and to detect defects~\cite{istqb2012standardGlossary}. This broad definition of testing comprises dynamic testing activities like classic unit or system testing, but also static testing activities, where artifacts are not executed, like static analysis or reviews. The tested software-based system is called \emph{system under test} (SUT). Traditional dynamic testing evaluates whether the SUT behaves as expected or not by executing a \emph{test suite}, i.e., a \emph{finite} set of test cases suitably \emph{selected} from the usually infinite execution domain ~\cite{IEEE2014SWEBOK}. In a generic sense, dynamic testing can also be defined as evaluating software by observing its execution~\cite{ammann2016introduction}, which may also subsume \emph{runtime monitoring} of live systems.
	
	After executing a test case, the observed and intended behaviors of a SUT are compared with each other, which then results in a \emph{verdict}~\cite{felderer2016security}. Verdicts can be either of \emph{pass} (behaviors conform), \emph{fail} (behaviors don't conform), and \emph{inconclusive} (not known whether behaviors conform)~\cite{ISO1994ITOpenSystemInterconnection}. A \emph{test oracle} is a mechanism for determining the verdict. The observed behavior may be checked against user or customer needs (commonly referred to as testing for \emph{validation}) or against a specification (testing for \emph{verification}). So the oracle compares an expected output value with the observed output. But this may not always be feasible, especially in the context of data-intensive software systems. For instance, consider data-intensive systems that produce complex output, like complicated numerical simulations or varying output based on a prediction model, where defining the correct output for a given input and then comparing it with the observed output may be non-trivial and error-prone. This problem is referred to as the \emph{oracle problem} and it is recognized as one of the fundamental challenges of software testing~\cite{weyuker1982testing}. \emph{Metamorphic testing}~\cite{segura2016survey} is a technique conceived to alleviate the oracle problem. It is based on the idea that often it is simpler to reason about relations between outputs of a program, than it is to fully understand or formalize its input-output behavior.
	
	A \emph{failure} is an undesired behavior. Failures are typically observed (by resulting in verdict fail) during the execution of the system under test. A \emph{fault} is the cause of a failure. It is a static defect in the software, usually caused by human error in the specification, design, or coding process. During testing, it is the execution of faults in the software that causes failures. Differing from active execution of test cases, passive testing only monitors running systems without interaction.
	
\begin{figure}[htb]
		\sidecaption
		\includegraphics[width=\textwidth]{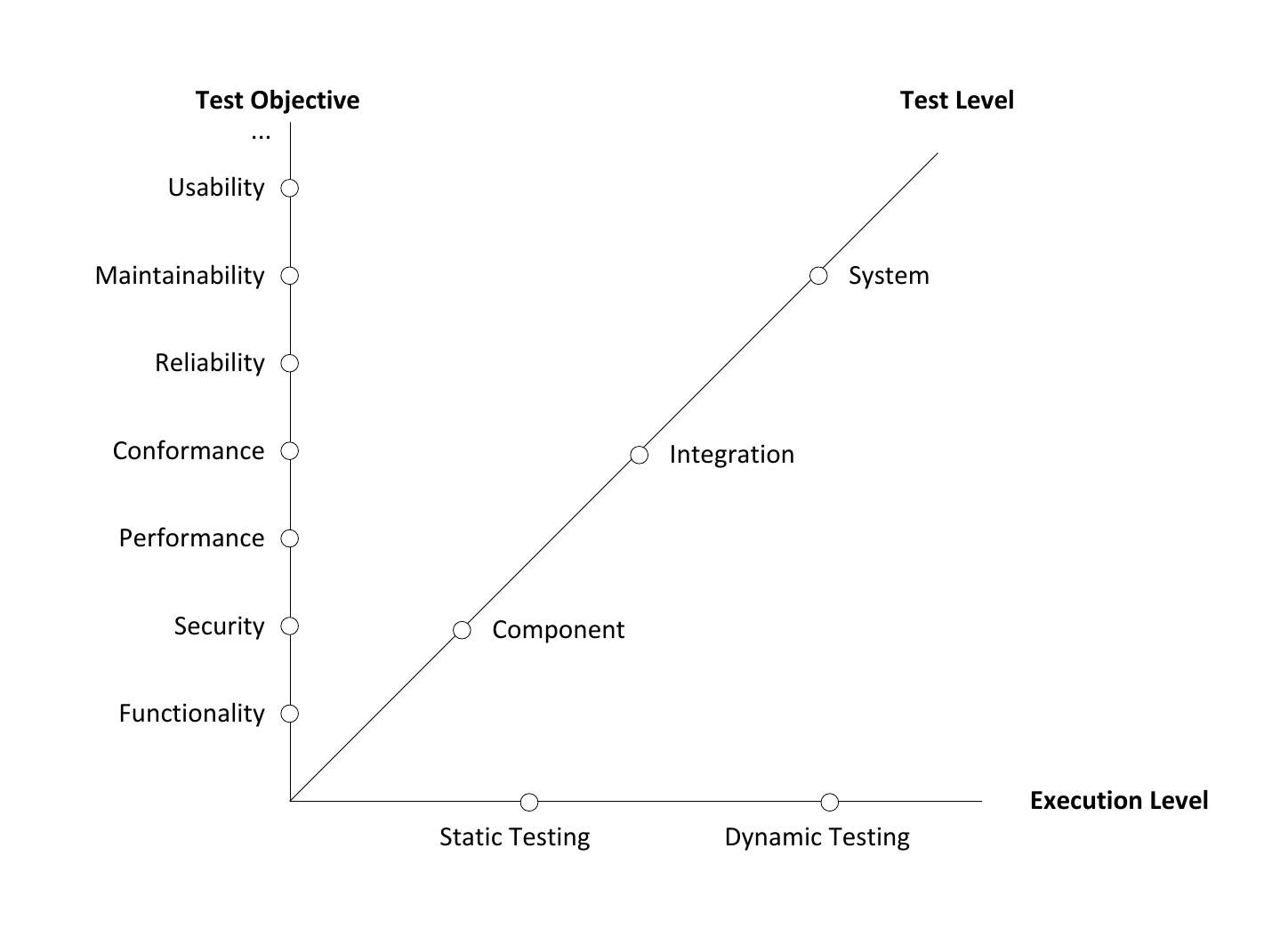}
		\caption{Testing dimensions: test level, test objective and execution level of traditional software testing}
		\label{fig:testingdims}
	\end{figure}

	Refining previous classifications~\cite{utting2007PracticalMBTAToolsApproach,felderer2016security}, testing can be classified utilizing the three dimensions: test objective, test level and execution level (see Fig.~\ref{fig:testingdims}).
	
	\emph{Test objectives} are reasons or purposes for designing and executing a test. The reason is either to check the functional behavior of the system or its non-functional properties. \emph{Functional testing} is concerned with assessing the functional behavior of an SUT, whereas \emph{nonfunctional testing} aims at assessing nonfunctional requirements with regard to quality characteristics like security, performance, conformance, reliability, maintainability or usability.
	
	The \emph{test level} addresses the granularity of the SUT and can be classified into component, integration and system testing. It also determines the \emph{test basis}, i.e., the artifacts to derive test cases. \emph{Component testing} (also referred to as \emph{unit testing}) checks the smallest testable component (e.g., a class in an object-oriented implementation or a single electronic control unit) in isolation. \emph{Integration testing} combines components with each other and tests those as a subsystem, that is, not yet a complete system. \emph{System testing} checks the complete system, including all subsystems. A specific type of system testing is \emph{acceptance testing} where it is checked whether a solution works for the users of a system. \emph{Regression testing} is a selective retesting to verify that modifications have not caused side effects and that the SUT still complies with the specified requirements~\cite{Radatz1990IEEEStandardGlossary}.
	
	The \emph{execution level} addresses whether and how the SUT is executed. \emph{Static testing} checks software development artifacts (e.g., requirements, design or code) without execution of these artifacts. \emph{Dynamic testing} actively executes a system under test and evaluates its results.
	
	The process of testing comprises the core activities test planning, design, automation, execution, and evaluation~\cite{istqb2012standardGlossary}. According to~\cite{istqb2012standardGlossary}, \emph{test planning} is the activity of establishing or updating a test plan. A test plan is a document describing the objectives, scope, execution levels, approaches, resources, and schedule of intended test activities. It identifies, among others, concrete objectives, the features to be tested, the test design techniques, and exit criteria to be used and the rationale of their choice. A test objective is a reason, which can either be to check the functional behavior of a system or its nonfunctional or (software) quality properties, for designing and executing a test. Exit criteria are conditions for permitting a process to be officially completed. They are used to report against and to plan when to stop testing. Adequacy criteria like coverage criteria aligned with the tested feature types and the applied test design techniques are typical exit criteria. Once the test plan has been established, test control begins. It is an ongoing activity in which the actual progress is compared against the plan which often results in concrete measures. 
	
	During the \emph{test design} phase the general testing objectives defined in the test plan are transformed into tangible test conditions and abstract test cases. In \emph{criteria-based test design}, one designs test cases that satisfy specific engineering goals such as coverage criteria. In \emph{human-based test design}, one designs test cases based on domain knowledge of the system and human knowledge of testing. \emph{Test automation} comprises tasks to make the abstract test cases executable. This includes tasks like preparing test harnesses and test data, providing logging support or writing test scripts which are necessary to enable the automated execution of test cases. In the \emph{test execution} phase, the test cases are then executed and all relevant details of the execution are logged and monitored. Finally, in the \emph{test evaluation} phase, the exit criteria are evaluated and the logged test results are summarized in a test report.

	\subsection{Data Quality}
	\label{subsec:dataquality}
	
	Quality, in general, has been defined as the totality of characteristics of a product that bear on its ability to satisfy stated or implied needs~\cite{iso2000qm}. This generic definition can be instantiated to software and data quality,  as capability of a software and data product, respectively, to satisfy stated and implied needs when used under specified conditions.  
	For software systems, according to ISO/IEC 25010~\cite{ISO2011Square}, these quality characteristics are: functional suitability, performance efficiency, compatibility, usability, reliability, security, maintainability and portability for product quality as well as effectiveness, efficiency, freedom from risk and context coverage for quality in use. 
	According to ISO/IEC 25012~\cite{ISO2008DataQuality} data quality characteristics in the context of software development can be classified into inherent and system-dependent data characteristics. 
	
	\emph{Inherent data quality} refers to the degree to which data quality characteristics  have the intrinsic potential to satisfy stated and implied needs when data is used under specified conditions. From the inherent point of view, data quality refers to data itself, in particular to data domain values and possible restrictions (e.g., business rules governing the quality required for the characteristic in a given application), relationships of data values (e.g., consistency) and meta-data. 
	
	\emph{System-dependent data quality} refers to the degree to which data quality is reached and preserved within a system when data is used under specified conditions. From this point of view, data quality depends on the technological domain in which data are used and it is achieved by the capabilities of systems' components such as hardware devices or sensors (e.g., to make data available or to obtain the required precision) as well as system and other software (e.g., backup software to achieve recoverability or data processing software).

	According to ISO/IEC 25012~\cite{ISO2008DataQuality}, inherent data quality characteristics are:
	\begin{itemize}
		\item \emph{Accuracy}, i.e., the degree to which data has attributes that correctly represent the true value of the intended attribute of a concept or event in a specific context of use.
		\item \emph{Completeness}, i.e., the degree to which data has attributes that are free from contradiction and are coherent with other data in a specific context of use. It can be either or both among data regarding one entity and across similar data for comparable entities.
		\item \emph{Credibility}, i.e., the degree to which data has attributes that are regarded as true and believable by users in a specific context of use. Credibility includes the concept of authenticity (the truthfulness of origins, attributions, commitments).
		\item \emph{Currentness}, i.e., the degree to which data has attributes that are of the right age in a specific context of use.
	\end{itemize}

	According to ISO/IEC 25012~\cite{ISO2008DataQuality}, inherent and system-dependent characteristics are :
	\begin{itemize}
		\item \emph{Accessibility}, i.e., the degree to which data can be accessed in a specific context of use, particularly by people who need supporting technology or special configuration because of some disability.
		\item \emph{Compliance}, i.e., the degree to which data has attributes that adhere to standards, conventions or regulations in force and similar rules relating to data quality in a specific context of use. 
		\item \emph{Confidentiality}, i.e., the degree to which data has attributes that ensure that it is only accessible and interpretable by authorized users in a specific context of use.
		\item \emph{Efficiency}, i.e., the degree to which data has attributes that can be processed and provide the expected levels of performance by using the appropriate amounts and types of resources in a specific context of use.
		\item \emph{Precision}, i.e., the degree to which data has attributes that are exact or that provide discrimination in a specific context of use.
		\item \emph{Traceability}, i.e., the degree to which data has attributes that provide an audit trail of access to the data and of any changes made to the data in a specific context of use.
		\item \emph{Understandability}, i.e., the degree to which data has attributes that enable it to be read and interpreted by users, and are expressed in appropriate languages, symbols and units in a specific context of use.
	\end{itemize}
	
	According to ISO/IEC 25012~\cite{ISO2008DataQuality}, system-dependent characteristics are :
	\begin{itemize}
		\item \emph{Availability}, i.e., the degree to which data has attributes that enable it to be retrieved by authorized users and/or applications in a specific context of use.
		\item \emph{Portability}, i.e., the degree to which data has attributes that enable it to be installed, replaced or moved from one system to another preserving the existing quality in a specific context of use.
		\item \emph{Recoverability}, i.e., the degree to which data has attributes that enable it to maintain and preserve a specified level of operations and quality, even in the event of failure, in a specific context of use.
	\end{itemize}
	
	Big data is a term used to refer to data sets that are too large or complex for traditional data-processing software to adequately process them. It is high-volume, high-velocity and high-variety information assets that demand cost-effective, innovative forms of information processing for enhanced insight and decision making.
	Big data can be described by the following characteristics:
	\begin{itemize}
		\item \emph{Volume} refers to the quantity of generated and stored data that determined the value and potential insight
		\item \emph{Velocity} refers to the speed at which the data is generated and processed to meet the demands and challenges
		\item \emph{Variety} refers to the different types of data that have to be processed, i.e., text, images, audio and video
		\item \emph{Veracity} refers to the refers to the data quality and the data value that can be measured based on the quality criteria highlighted before
	\end{itemize}

	\section{Data-Intensive Software Systems}
	\label{sec:diss}
	
	Systems that process large volumes of data have commonly been referred to as "data-intensive". However, the amount of data that is considered as large is relative and changes over time. Thus, information systems that stored and retrieved large volumes of data at their time, may be no longer considered as "data-intensive" today. This made the concept of "data-intensive" systems relative to time and recent developments in storage capacities and processing capabilities.
	
	Today, data-intensive systems do not only process large volumes of data, but differ from other systems with respect to the following aspects:
	\begin{itemize}
		\item The data is stored and retrieved but depending on the context also collected, generated, manipulated and redistributed (i.e., processed).
		\item The data is "big" data and fulfills the characteristics described in the previous section (i.e., volume, velocity, variety, and veracity).
		\item The data does influence in addition to the operation phase also the analysis, design, implementation, testing, and maintenance phase of the system lifecylce.
	\end{itemize}
	
	These data-intensive systems are more than large data stores that allow the storage and retrieval of large volumes of data. Data-intensive systems additionally comprise the collection of data from different sources, the generation of new data, the manipulation of existing data and the redistribution of data to data consumers.
	
	The data itself fulfills the characteristics of big data. The different data sources that are used by data-intensive systems provide large volumes of data (volume) in varying formats (variety). Furthermore, the data is processed timely to meet the demands of its users (velocity). Finally, the high dependency on data requires the maintenance of high data quality (veracity) throughout the system.
	
	With respect to the central role that data has in all phases of the system lifecycle (with the consequence that the system behavior is determined by the data), data-intensive systems can be opposed to specification-driven systems where the behavior is explicitly specified and coded in (deterministic) algorithms.

	To summarize, data-intensive systems are defined as follows:\\ 
	
	\emph{Data-intensive systems are systems that may handle, generate, and process a large volume of data of different nature and from different sources over time orchestrated by means of different technologies and with the goal to extract value from data for different types of businesses. They pose specific challenges in all phases of the system lifecycle and over time they might evolve to very large, complex systems as data, technologies and value evolve.}\\
	
	In the following sections, the architecture and quality assurance of data-intensive systems are discussed.

	
	
	\subsection{Architecture of Data-Intensive Systems} \label{subsection:architecture}

	Recent years have seen the rapid growth of interest on developing data-intensive systems. These systems include technologies like Hadoop/MapReduce, NoSQL, or systems for stream processing~\cite{casale2015dice}.
	Specific technologies like databases, caches, search indexes, stream processing, or batch processing enable the development of data-intensive systems. They provide services like decision support, system behavior monitoring or reporting applicable in different business domains.
	Thus, a data-intensive system is typically built from standard building blocks that provide commonly needed functionality. For example, many systems need databases, search indexes, or batch processing.  
	As there are various options for database systems, or approaches to caching, the most appropriate technologies for the respective system have to be chosen among the numerous possibilities. Key quality attributes for these decisions are, according to Kleppmann~\cite{kleppmann2017designing}, reliability, scalability, and maintainability. 
	
\begin{figure}[htb]
		\sidecaption
		\includegraphics[width=\textwidth]{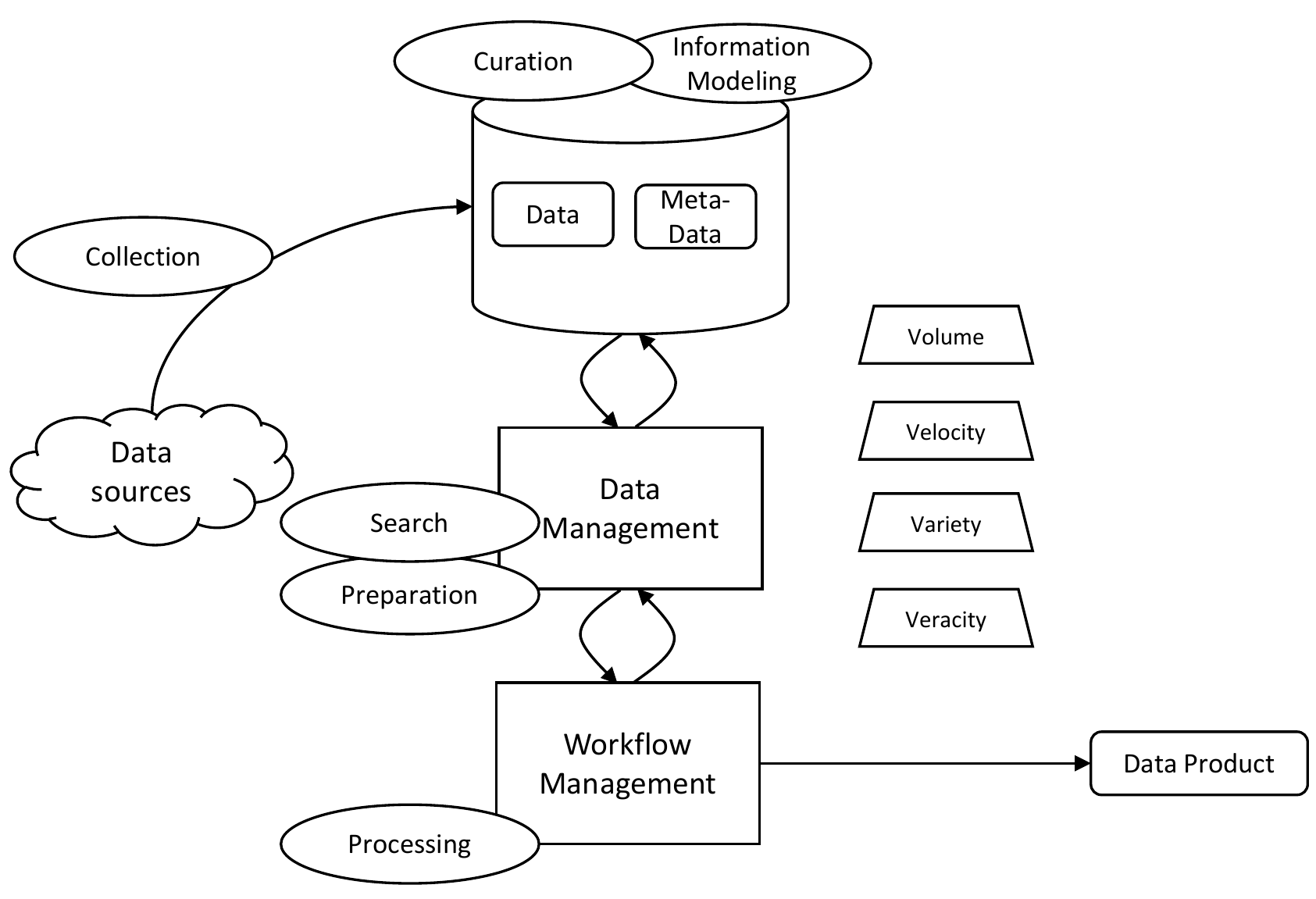}
		\caption{Conceptual structure of a data-intensive system.}
		\label{fig:diss}
	\end{figure}

	The building blocks that assemble a data-intensive system are visualized in the conceptual structure of a data-intensive system given in Fig.~\ref{fig:diss}, which is based on the system lifecycle of data-intensive systems presented in~\cite{mattmann2011architecting}. Following the path of the data through the system, first the data is collected from the different data sources with possible varying protocols (e.g., ftp, http) and data formats like domain-specific exchange formats ("Data sources"). 
	Next, in the staging area, it is ensured that the data conforms to an unified interpretation, sharing and preservation (curation). This includes describing the data with meta-data to ease further processing. The meta-data are created, organized and classified according to the information modeling of the data-intensive system.
	Thereafter, the data and its meta-data are stored into an archive system ("Data management"). This part of the system supports searching on the data and further prepares the data for its later usage in the workflow management component.
	Following the data management, the workflow management is responsible for processing the data. It includes tasks to process the data (e.g., calculations) and workflows that align tasks into a processing pipeline that can make data-based decisions on which concrete tasks to execute. 
	In the last step, the data leaves the data-intensive system as part of a data product. Examples therefore are reports, prediction models, analyses and recommendations or generated data. It represents the data output of the data-intensive system that may be further processed within a system.
	The four big data characteristics (see Section \ref{subsec:dataquality}) that are next to the system components highlight that volume, velocity, variety and veracity are inherent in every part of the system. Each component, and activity (e.g., collection, curation or processing) have to take these characteristics into account. This reflects the intensive dependency of these systems on data.
	
	
	
	\subsection{Quality Assurance of Data-Intensive System}\label{sec:qa-dis}
	
	Data-intensive systems process large volumes of data and provide through the processing of it high value to the business.
	Thus, failures and other quality issues in data-intensive systems are incredibly costly (e.g., production failures), because of their affect on the business. As a result, new types of quality assurance activities and concerns raise. For example, data debugging i.e., discovering system failures caused by well-formed but incorrect data is a primary issue in the quality assurance of data-intensive systems.
	
	Hummel et al.~\cite{hummel2018collection} identified eight challenges of quality assurance in the context of data-intensive systems.
	\begin{itemize}
		\item \emph{Challenging Visualization and Explainability of Results}. Data needs to be visualized with the right balance between data dimensions and resolution, in order to support the user to understand and assess the validity of the data. Furthermore, the processing that lead to a particular result is difficult to explain (e.g., in deep learning). Thus, trustworthiness and understandability are important challenges for data-intensive systems.
		\item \emph{Non-Intuitive Notion of Consistency}. The large volume of data that is processed by data-intensive systems requires to weaken the notion of data consistency for performance reasons. These inconsistencies are confusing for users not used to softened consistency notions.
		\item \emph{Complex Data Processing and Different Notions of Correctness}. The numerous processing steps and interconnections between data makes the processing of data in data-intensive systems complex. The notion of correctness becomes difficult to define, which complicates the testability of such systems.
		\item \emph{High Hardware Requirements for Testing}. Data-intensive systems are expected to process large volumes of data. Thus, testing requires to consider issues like performance or scalability that are related to processing of big data. However, this requires hardware similar to its later application environment, which may be not possible because of involved operating costs of such environments.
		\item \emph{Difficult Generation of Adequate, High-Quality Data}. The data provided for testing should represent realistic and application-specific data in order to meaningfully test a data-intensive system. Thus, all big data characteristics described in Section \ref{subsec:dataquality} should be covered, which makes test data generation and management a challenging task, especially when also taking data quality aspects into account.
		\item \emph{Lack of Debugging, Logging, and Error-Tracing Methods}. The architecture of data-intensive systems results into a distributed system that complicates debugging. Furthermore, logging is scattered over the different components of a system. As a result, tracing errors back to their origin or understanding system behavior based on the logs become inherently difficult.
		\item \emph{State Explosion in Verification}. The processing of large volumes of data that requires distributed computing to process its requests, results into an exponential state explosion that makes verification approaches complex to apply.
		\item \emph{Ensuring Data Quality}. The data quality of the large volumes of data that are collected, processed and aggregated by data-intensive systems is difficult to assess computationally and semantically.
	\end{itemize}
	
	\section{Cyber-Physical Production Systems as Data-Intensive Systems} 
	\label{sec:lcps}
	
\begin{figure}[htb]
		\centering
		\includegraphics[width=\textwidth]{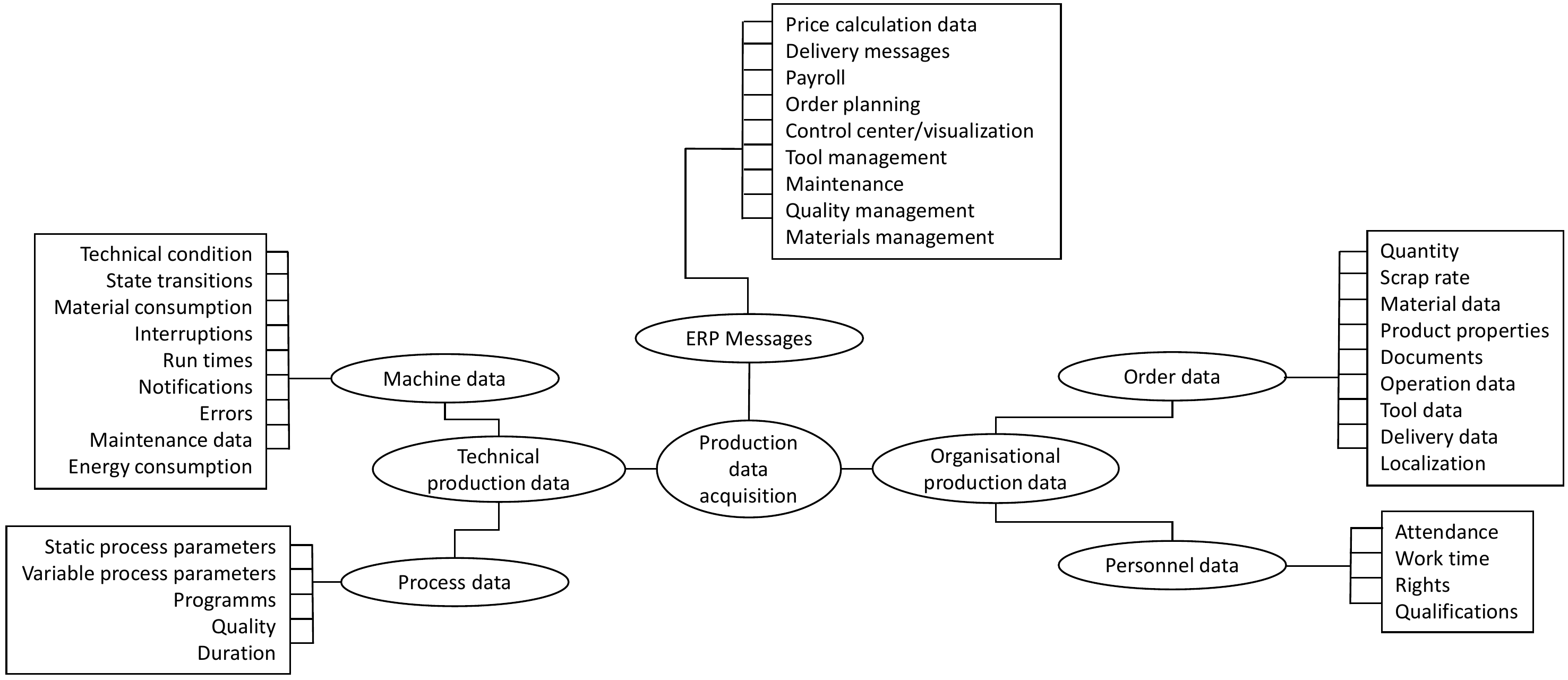}
		\caption{Overview of data sources in the context of production systems.}
		\label{fig:data-production-system}
	\end{figure}

	Industrial production systems like robots, steel mills or manufacturing plants are large and long-running cyber-physical systems that generate and process large volumes of data from various sources (see Fig. \ref{fig:data-production-system}) and types (e.g., order data, personnel data or machine data). Data sources of typical production systems are among others technical production data (like machine or process data), organizational production data (like order or personnel data) and messages from the ERP system. Thus, cyber-physical production systems represent data-intensive systems that are of high value to the business and require testing. The production is tightly coupled to the data-intensive systems, which puts high requirements on the standard of quality assurance of the system and the data it operates on.
	
	The architecture of a large cyber-physical production system as used in manufacturing plant is conceptually consistent with the architecture described in Section \ref{subsection:architecture}. Data of various sources (e.g., sensors, orders, processes) is collected, unified and organized according to the later information needs of the production system (e.g., order status). Employees or intelligent systems (as envisioned by Industry 4.0~\cite{wang2016implementing,foidl2015research}) can search within the data and extract relevant information for their specific data needs. Finally, every production system has some kind of workflow management that processes the data in order to generate data products like aggregated reports or prediction models (e.g., for predictive maintenance) to support decision making.
	
	Production failures or unintended production stops are possible results of software faults in the cyber-physical production system and consequence of inappropriate testing. A cyber-physical production system in a manufacturing plant, which represents a complex and large production system that processes large volumes of data, is a critical business asset that requires careful testing to mitigate production related risks. Furthermore, the software and data quality of the system has a strong affect on the production itself. Higher quality standards can lead to optimizations and improvements of the overall production, whereas quality problems can lead to a decline in production. Challenges to the quality assurance (see Section \ref{sec:qa-dis}) like high hardware requirements for testing (e.g., related to robots or production line) or the difficulty to ensure the quality of the data from various sources (e.g., machines, orders, processes) are specifically present in production systems. Moreover, the trend to smart factories, that, for example, encourage the intelligent communication between robots, further increases the requirements on quality assurance. As a result, the research on testing of data-intensive software systems (see Section \ref{sec:testingdiss} and Section \ref{sec:discussion}) is of great importance for large cyber-physical production systems in general and their quality assurance in particular.

	\section{Testing Data-Intensive Software Systems}
	\label{sec:testingdiss}
	
	Research on testing data-intensive systems is becoming of high academic and practical importance. Only few recent papers (e.g., Hummel et al.~\cite{hummel2018collection} as presented in Section~\ref{sec:qa-dis}) discussed challenges and derived open questions to start guiding research in that field. However, primary studies contributing approaches on testing data-intensive systems are still rare. In this section, we report the result of our literature study on existing research on testing data-intensive systems. The study is not intended to be systematic, but rather exploratory to get an initial insight on the aspects of the research that are currently under investigation.      
	
	\subsection{Literature Study on Testing Data-Intensive Systems}
	\label{sec:literature}
	
	We performed an exploratory and preliminary literature study to get initial insights into the research on testing data-intensive systems. Papers have been collected from Google Scholar\footnote{\url{http://scholar.google.at}} and the IEEE Xplore Digital Library\footnote{\url{https://ieeexplore.ieee.org}} by searching with keywords "Data-Intensive System" (OR "Data-Intensive Systems") AND "Software Testing". Included papers are written in English and are journal articles, conference or workshop papers, book chapters, technical reports or thesis works that cover at least one testing activity as well as big data or data processing aspects. The search initially returned 133 papers. After reading title and abstract and eventually the body of the work, papers were removed if they: 
	\begin{itemize}
		\item Misused the search terms (e.g., "Data-Oriented Systems" as "Data-Intensive System"), 
		\item Only used the search term "Data-Intensive System" or "Software Testing" in their citations, 
		\item Only referred to data-intensive systems as one of the systems under analysis with no specific discussion or exploitation of the characteristics of data-intensive systems, or 
		\item Mention "Data-Intensive System" only in the related work section or as an example context in which an approach can eventually work. 
	\end{itemize}
	Taking these criteria into account, we could finally include 16 relevant papers. This indicates that software testing of data-intensive systems is not yet an established and clearly-defined term. The 16 papers were further classified according to three major areas, (i.e., Testing, System, Data), related categories (e.g., Level of Testing), and factors (e.g., Component, Integration, and System for Level of Testing) based on the following criteria:
	\begin{itemize}
		\item Testing
		\begin{itemize}
			\item Level of Testing (Component, Integration, System)
			\item Test Activity (Test Management, Test-Case Design (Criteria-based), Test-Case Design (Human Knowledge-Based), Test Automation, Test Execution, Static Analysis, Test Evaluation (Oracle))
		\end{itemize}
		\item System
		\begin{itemize}
			\item System Quality (Security, Performance, Conformance, Usability etc.)
			\item System Artifact (SQL Query, System Code, User Form etc.)
		\end{itemize}
		\item Data
		\begin{itemize}
			\item Big Data Aspect (Volume, Velocity, Variety, Veracity, Value)
			\item Data Processing Aspect (Collection, Generation, Manipulation, Redistribution)
			\item Data Quality Aspect (Integrity etc.)
			\item Data Technology (Database, Cache etc.)
		\end{itemize}
	\end{itemize}
	
	The applied classification scheme was derived from established classifications as presented in Section~\ref{sec:background} and Section~\ref{sec:diss}, experience of the authors and an analysis of relevant papers. Open coding (to identify the categories) and axial coding (to relate categories) of grounded theory~\cite{corbin1990grounded} on titles and abstracts of such papers have been used to verify and refine the initial categorization.
	
	\begin{figure}[htb]
		\centering
		\includegraphics[scale=0.85,trim={60 0 80 0},clip]{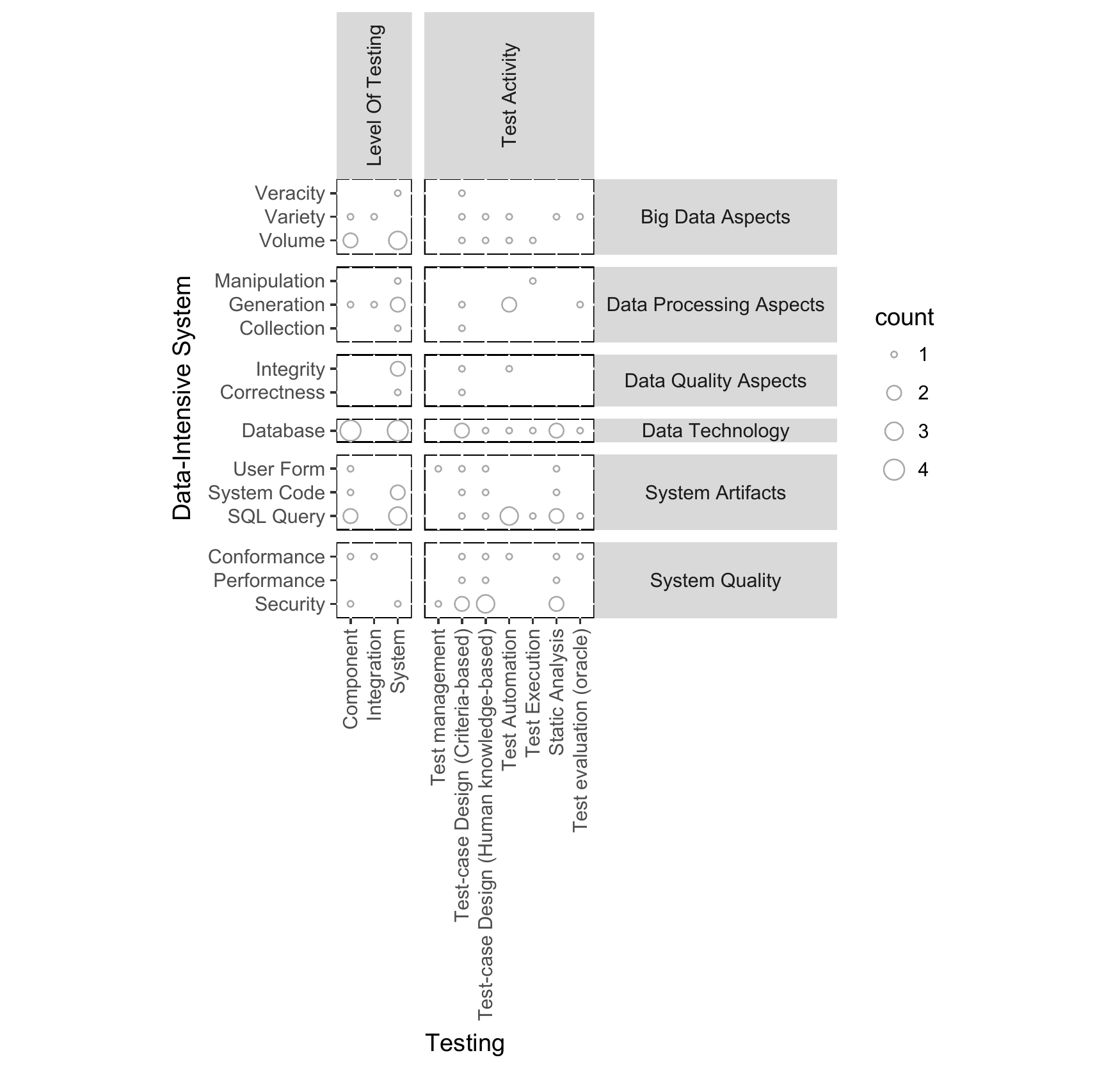}
		\caption{Classification of relevant papers.}
		\label{fig:myBubbles}
	\end{figure}
	\subsection{Classification of Papers}
	\label{sec:classification}
	
	The 16 relevant papers were classified according to the categories and factors defined in Section~\ref{sec:literature}. The classification has been performed by reading the body of each paper. Papers may be classified into multiple factors of the same category or multiple categories of the same area.
	
	Figure~\ref{fig:myBubbles} illustrates the resulting classification of the included papers by comparing categories and factors characterizing software testing with the ones of data-intensive system (i.e., categories classified as \emph{System} and \emph{Data} in Section \ref{sec:literature}). 
	One may immediately recognize that not all factors are included in the figure. For instance, Velocity for the Big Data aspects or Cache for the Data Technology aspects are never considered when it comes to software testing in the investigated set of papers. For each factor, the maximum count is four (i.e., a quarter of the investigated papers), which is attained for (1) Database in the Data Technology category at Component and System Level of Testing, (2) Test-Case Design (Human Knowledge-Based) and Test Automation in the Test Activity category for what concerns SQL Query in System Artifacts and at the System Level Testing and Security in System Quality, respectively, (3) Volume in Big Data aspects concerning System Level of Testing.

	\section{Discussion} 
	\label{sec:discussion}
	
	In this section, we discuss the current state of testing data-intensive software systems and present directions of research that have the potential to significantly contribute to solving key challenges of testing data-intensive software systems.
	
	The initial literature survey presented in the last section revealed that the automatic search in digital libraries returns a large number of false positives (i.e., 117 out of 133 papers) where the terms of the search are used only marginally, often to mention only future application or related work of a proposed approach. However, we also need to acknowledge that our approach is just exploratory and may have missed some relevant papers. In particular, we decided not to perform snowballing from the papers retrieved from the digital libraries as research on data-intensive system is rather new and backward snowballing seemed not to be promising after a first check of the references of the retrieved papers.
	With the classification of the relevant papers, we found that some aspects that commonly refer to testing or data-intensive software system are not yet explored when it comes to testing of data-intensive systems. For instance, for velocity as a big data aspect suitable testing approaches for data-intensive systems are not available. Velocity is of paramount importance in modern data-intensive systems, especially in the context of real time analytics, where results are delivered in a continuous fashion and results are only useful if the delay is very short. However, before focusing on testing for velocity, the research community is still struggling with the uncertainty of the output of test cases and the definition of suitable oracles (\cite{deBayserEtAl2015}) and the volume and variety of data that data-intensive systems can potentially handle (e.g., \cite{RussoEtAl2015}, \cite{GadlerEtAl2017}). A certain number of papers concern testing the correctness and the integrity of the data stored in databases by data-intensive systems. For instance, static analysis has been used to test SQL statements embedded in the code of data-intensive systems to ensure database integrity \cite{Nagy2013}. 
	
	However, there are several promising directions of research and new frontiers that are highly relevant in the context of data-intensive systems that have not yet been fully exploited.
	
	Testing of data-intensive software systems is problematic because these systems belong to the class of "non-testable" software, where typically no test oracles exist for verification~\cite{otero2015research}. A proven approach for testing such systems is metamorphic testing~\cite{segura2016survey}, which requires the discovery of metamorphic relations that can be used to test the validity of machine learning algorithms. For instance, Xie et al.~\cite{xie2011testing} provide an approach to metamorphic testing of machine-learning-based classification algorithms. The authors enumerate the metamorphic relations that classifiers would be expected to demonstrate and then determine for a given implementation whether each relation is a necessary property for the corresponding classification algorithm. If this is the case, then failure to exhibit the relation indicates a fault.
	
	Testing of machine learning algorithms, especially for sophisticated algorithms like deep learning systems, requires specific testing approaches. To that end, recently several adequacy criteria for deep learning systems like neuron coverage~\cite{pei2017deepxplore}, surprise adequacy~\cite{kim2018guiding} or criteria derived from modeling deep learning systems as abstract state transition systems~\cite{du2018deepcruiser} were defined. 
	However, testing algorithms is not sufficient as the integration of algorithms into systems can be complex, leading to problems and defects being injected along the way. Specific concerns mentioned in the literature include untrustworthy data, bad assumptions, and incorrect mathematics among others~\cite{shull2013getting}. Tian et al.~\cite{tian2018deeptest} propose an approach based on metamorphic testing to test deep learning systems in the context of autonomous driving. As especially the environment is very complex, simulation also plays an important role for testing data-intensive systems as used in the context of autonomous driving, where large, complex, and independent system interact and form so called Systems-of-Systems~\cite{dersin2014systems}. 
	
	One step further, testing can even be shifted to the running system. This novel approach to testing of data-intensive systems performs monitoring of the runtime environment to assess the behavior and changes in the system. System changes under test are exposed to a limited number of users. The measured runtime data generated through their interaction with the changes are compared to users that were not exposed to the changes. The analysis of the collected experimentation data reveals the influence of the chance on predefined characteristics like performance or usability. The approach of system change assessment is also continuous experimentation (\cite{fagerholm2014building}) and promising especially also in the context of testing data-intensive systems~\cite{auer2018current}.
	
	Observing unexpected execution behavior is also used to build self-assessment oracles, a new type of oracles introduced in the anticipatory testing approach \cite{Tonella2018}. Such an approach aims at detecting failures before they even occur. Its oracles can be crucial for data-intensive systems where the volume of the input and the uncertainty of the output are key challenges in designing test cases~\cite{hummel2018collection}.  
	
\begin{figure}[htb]
		\sidecaption
		\includegraphics[width=\textwidth]{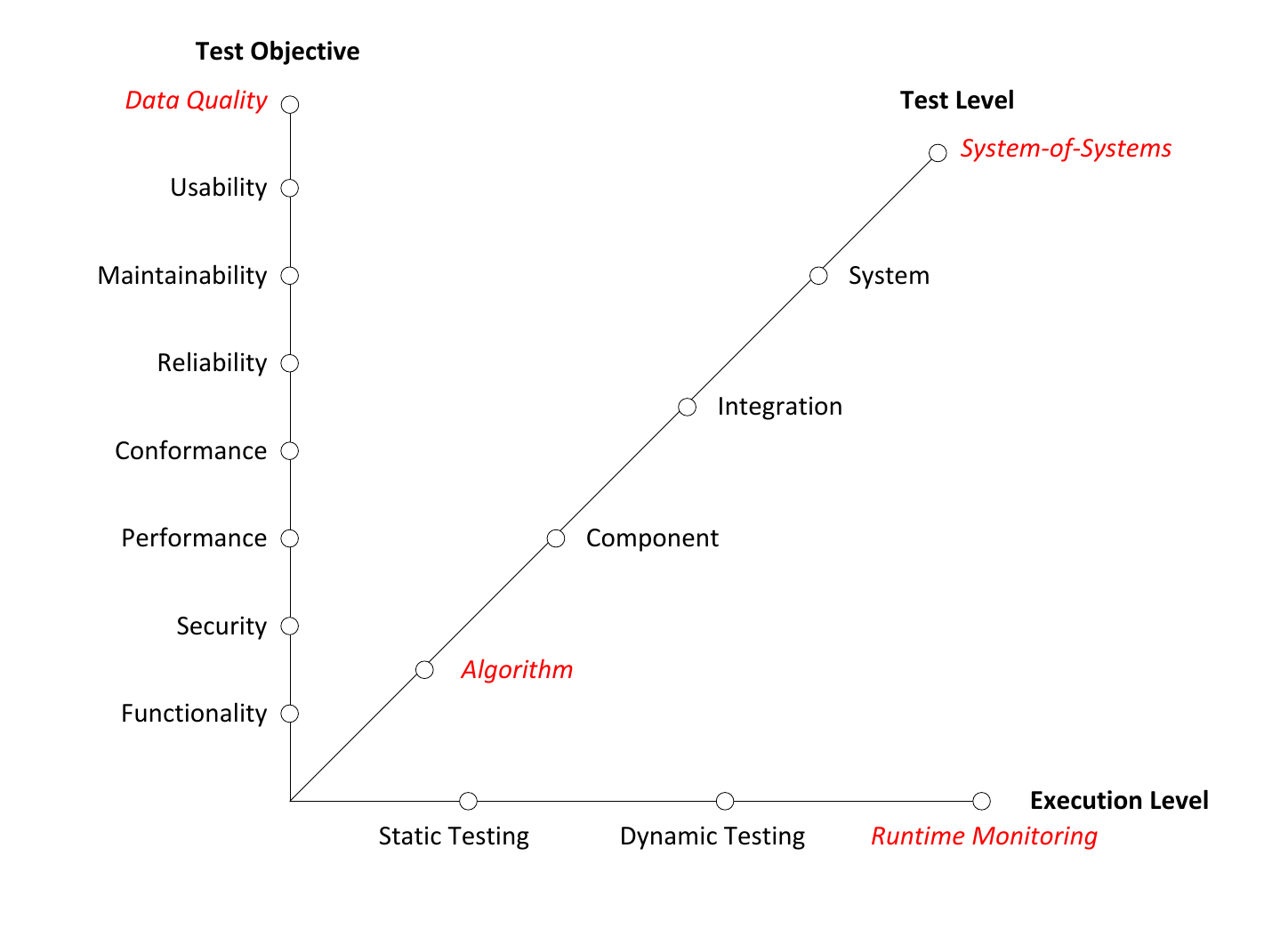}
		\caption{Testing dimensions test level, test objective and execution level of testing data-intensive software systems}
		\label{fig:testingdims-diss}
	\end{figure}

	Taking the discussion into account, the categories on the testing dimensions for traditional systems shown in Fig.~\ref{fig:testingdims} have to be extended for testing data-intensive software systems. The testing dimensions and their values for data-intensive software systems are shown in Fig.~\ref{fig:testingdims-diss}. The added categories are shown in italic.
	
	The dimension test level is extended by the values \emph{Algorithm} and \emph{System-of-Systems}. On the one hand, the testing of algorithms in isolation becomes essential (e.g., for classification algorithms or deep learning algorithms). On the other hand, also \emph{Systems-of-Systems}, for instance in the context of autonomous driving, require new testing approaches where especially simulation of the environment plays an important role. 
	The dimension execution level is extended by \emph{Runtime Monitoring}, where a running system is passively observed to test changes or unspecified system behavior, in contrast to active penetration in case of dynamic testing.
	Finally, the dimension test objective is extended by \emph{Data Quality}, which comprises testing of data and big data quality properties like correctness or velocity.
	
	\section{Conclusion}
	\label{sec:conclusion}
	
	Software engineering research is needed to fully understand the role of testing in data-intensive software systems (like large cyber-physical production systems) and to provide approaches and frameworks to properly address this challenging and critical issue. The goal of this chapter is to contribute into this direction by providing basic terminology, an overview of the literature, a discussion, and future directions of research.
	
	In this chapter, we first presented the background as well as the basic terminology and then we discuss the current state and new frontiers on testing data-intensive software systems that are also relevant for testing cyber-physical production systems. To this end, we first provided a definition of data-intensive system (Section \ref{sec:diss}) that clearly distinguishes them from other types of software systems and characterized cyber-physical production systems as data-intensive systems. The definition also helped us to derive the major factors characterizing such systems and drive the investigation on the existing research (Section \ref{sec:classification}). We further discussed the current state and new frontiers on testing data-intensive software systems (Section \ref{sec:discussion}). The frontiers include metamorphic testing, algorithm testing, testing systems-of-systems, data quality aspects, runtime monitoring, and anticipatory testing. Finally, we extended the testing dimensions for data-intensive software systems accordingly.
	
	\bibliographystyle{spmpsci}
	
	\bibliography{references}

\begin{thebibliography}{10}
\providecommand{\url}[1]{{#1}}
\providecommand{\urlprefix}{URL }
\expandafter\ifx\csname urlstyle\endcsname\relax
  \providecommand{\doi}[1]{DOI~\discretionary{}{}{}#1}\else
  \providecommand{\doi}{DOI~\discretionary{}{}{}\begingroup
  \urlstyle{rm}\Url}\fi

\bibitem{ammann2016introduction}
Ammann, P., Offutt, J.: Introduction to software testing.
\newblock Cambridge University Press (2016)

\bibitem{auer2018current}
Auer, F., Felderer, M.: Current state of research on continuous
  experimentation: a systematic mapping study.
\newblock In: 2018 44th Euromicro Conference on Software Engineering and
  Advanced Applications (SEAA), pp. 335--344. IEEE (2018)

\bibitem{deBayserEtAl2015}
de~Bayser, M., Azevedo, L.G., Cerqueira, R.: Researchops: The case for devops
  in scientific applications.
\newblock In: 2015 IFIP/IEEE International Symposium on Integrated Network
  Management (IM), pp. 1398--1404 (2015)

\bibitem{IEEE2014SWEBOK}
Bourque, P., Dupuis, R. (eds.): {G}uide to the {S}oftware {E}ngineering {B}ody
  of {K}nowledge {V}ersion 3.0 {SWEBOK}.
\newblock {IEEE} (2014).
\newblock Http://www.computer.org/web/swebok

\bibitem{casale2015dice}
Casale, G., Ardagna, D., Artac, M., Barbier, F., Nitto, E.D., Henry, A.,
  Iuhasz, G., Joubert, C., Merseguer, J., Munteanu, V.I., et~al.: Dice:
  quality-driven development of data-intensive cloud applications.
\newblock In: Proceedings of the Seventh International Workshop on Modeling in
  Software Engineering, pp. 78--83. IEEE Press (2015)

\bibitem{corbin1990grounded}
Corbin, J.M., Strauss, A.: Grounded theory research: Procedures, canons, and
  evaluative criteria.
\newblock Qualitative sociology \textbf{13}(1), 3--21 (1990)

\bibitem{dersin2014systems}
Dersin, P.: Systems of systems.
\newblock IEEE-Reliability Society. Technical Committee on Systems of Systems
  (2014)

\bibitem{du2018deepcruiser}
Du, X., Xie, X., Li, Y., Ma, L., Zhao, J., Liu, Y.: Deepcruiser: Automated
  guided testing for stateful deep learning systems.
\newblock arXiv preprint arXiv:1812.05339  (2018)

\bibitem{fagerholm2014building}
Fagerholm, F., Guinea, A.S., M{\"a}enp{\"a}{\"a}, H., M{\"u}nch, J.: Building
  blocks for continuous experimentation.
\newblock In: Proceedings of the 1st international workshop on rapid continuous
  software engineering, pp. 26--35. ACM (2014)

\bibitem{felderer2016security}
Felderer, M., B{\"u}chler, M., Johns, M., Brucker, A.D., Breu, R., Pretschner,
  A.: Security testing: A survey.
\newblock In: Advances in Computers, vol. 101, pp. 1--51. Elsevier (2016)

\bibitem{foidl2015research}
Foidl, H., Felderer, M.: Research challenges of industry 4.0 for quality
  management.
\newblock In: International Conference on Enterprise Resource Planning Systems,
  pp. 121--137. Springer (2015)

\bibitem{GadlerEtAl2017}
Gadler, D., Mairegger, M., Janes, A., Russo, B.: Mining logs to model the use
  of a system.
\newblock In: 2017 {ACM/IEEE} International Symposium on Empirical Software
  Engineering and Measurement, {ESEM} 2017, Toronto, ON, Canada, November 9-10,
  2017, pp. 334--343 (2017)

\bibitem{hummel2018collection}
Hummel, O., Eichelberger, H., Giloj, A., Werle, D., Schmid, K.: A collection of
  software engineering challenges for big data system development.
\newblock In: 2018 44th Euromicro Conference on Software Engineering and
  Advanced Applications (SEAA), pp. 362--369. IEEE (2018)

\bibitem{iso2000qm}
{ISO}: Iso 8402:1994 quality management and quality assurance -- vocabulary.
\newblock Tech. rep., ISO (1994)

\bibitem{ISO1994ITOpenSystemInterconnection}
ISO/IEC: {Information technology -- open systems interconnection -- conformance
  testing methodology and framework} (1994).
\newblock International ISO/IEC multi--part standard No. 9646

\bibitem{ISO2008DataQuality}
ISO/IEC: Iso/iec 25012:2008 software engineering -- software product quality
  requirements and evaluation (square) -- data quality model.
\newblock Tech. rep., ISO (2008)

\bibitem{ISO2011Square}
ISO/IEC: Iso/iec 25010:2011 systems and software engineering -- systems and
  software quality requirements and evaluation (square) -- system and software
  quality models.
\newblock Tech. rep., ISO (2011)

\bibitem{istqb2012standardGlossary}
{ISTQB}: Standard glossary of terms used in software testing. version 2.2.
\newblock Tech. rep., ISTQB (2012)

\bibitem{kim2018guiding}
Kim, J., Feldt, R., Yoo, S.: Guiding deep learning system testing using
  surprise adequacy.
\newblock arXiv preprint arXiv:1808.08444  (2018)

\bibitem{kleppmann2017designing}
Kleppmann, M.: Designing data-intensive applications: The big ideas behind
  reliable, scalable, and maintainable systems.
\newblock O'Reilly (2017)

\bibitem{mattmann2011architecting}
Mattmann, C.A., Crichton, D.J., Hart, A.F., Goodale, C., Hughes, J.S., Kelly,
  S., Cinquini, L., Painter, T.H., Lazio, J., Waliser, D., et~al.: Architecting
  data-intensive software systems.
\newblock In: Handbook of Data Intensive Computing, pp. 25--57. Springer (2011)

\bibitem{Nagy2013}
Nagy, C.: Static analysis of data-intensive applications.
\newblock In: 2013 17th European Conference on Software Maintenance and
  Reengineering, pp. 435--438 (2013)

\bibitem{otero2015research}
Otero, C.E., Peter, A.: Research directions for engineering big data analytics
  software.
\newblock IEEE Intelligent Systems \textbf{30}(1), 13--19 (2015)

\bibitem{pei2017deepxplore}
Pei, K., Cao, Y., Yang, J., Jana, S.: Deepxplore: Automated whitebox testing of
  deep learning systems.
\newblock In: Proceedings of the 26th Symposium on Operating Systems
  Principles, pp. 1--18. ACM (2017)

\bibitem{Radatz1990IEEEStandardGlossary}
Radatz, J., Geraci, A., Katki, F.: {IEEE} standard glossary of software
  engineering terminology.
\newblock Tech. rep., IEEE (1990)

\bibitem{RussoEtAl2015}
Russo, B., Succi, G., Pedrycz, W.: Mining system logs to learn error
  predictors: a case study of a telemetry system.
\newblock Empirical Software Engineering \textbf{20}(4), 879--927 (2015)

\bibitem{segura2016survey}
Segura, S., Fraser, G., Sanchez, A.B., Ruiz-Cort{\'e}s, A.: A survey on
  metamorphic testing.
\newblock IEEE Transactions on software engineering \textbf{42}(9), 805--824
  (2016)

\bibitem{shull2013getting}
Shull, F.: Getting an intuition for big data.
\newblock IEEE software \textbf{30}(4), 3--6 (2013)

\bibitem{tian2018deeptest}
Tian, Y., Pei, K., Jana, S., Ray, B.: Deeptest: Automated testing of
  deep-neural-network-driven autonomous cars.
\newblock In: Proceedings of the 40th International Conference on Software
  Engineering, pp. 303--314. ACM (2018)

\bibitem{Tonella2018}
Tonella, P.: 2019 -2023 ERC project: Precrime: Self-assessment Oracles for
  Anticipatory Testing (2018 - Accessed January 7th, 2019).
\newblock \urlprefix\url{www.pre-crime.eu}

\bibitem{utting2007PracticalMBTAToolsApproach}
Utting, M., Legeard, B.: {Practical Model-Based Testing: A Tools Approach}.
\newblock Morgan Kaufmann Publishers Inc., San Francisco, CA, USA (2007)

\bibitem{wang2016implementing}
Wang, S., Wan, J., Li, D., Zhang, C.: Implementing smart factory of industrie
  4.0: an outlook.
\newblock International Journal of Distributed Sensor Networks \textbf{12}(1),
  3159805 (2016)

\bibitem{weyuker1982testing}
Weyuker, E.J.: On testing non-testable programs.
\newblock The Computer Journal \textbf{25}(4), 465--470 (1982)

\bibitem{xie2011testing}
Xie, X., Ho, J.W., Murphy, C., Kaiser, G., Xu, B., Chen, T.Y.: Testing and
  validating machine learning classifiers by metamorphic testing.
\newblock Journal of Systems and Software \textbf{84}(4), 544--558 (2011)

\end{thebibliography}
\end{document}